\documentstyle[twocolumn,aps,floats,graphicx]{revtex}

\begin{document}

\newcommand{\bec}{\begin{center}}
\newcommand{\ec}{\end{center}}
\newcommand{\be}{\begin{equation}}
\newcommand{\ee}{\end{equation}}
\newcommand{\beqn}{\begin{eqnarray}}
\newcommand{\eeqn}{\end{eqnarray}}
\newcommand{\bet}{\begin{table}}
\newcommand{\ent}{\end{table}}
\newcommand{\bib}{\bibitem}

\wideabs{

\title{
Condensation energy, charge ordering and inter-layer coupling in cuprates 
}

\author{P. S\"ule} 
  \address{Research Institute for Technical Physics and Material Science,\\
Konkoly Thege u. 29-33, Budapest, Hungary,\\
sule@mfa.kfki.hu
}

\date{\today}

\begin{abstract}

\maketitle

The correlation between the condensation energy and the critical temperature is studied
within a charge ordered superlattice bilayer model
in which pairing is supported by inter-layer Coulomb energy gain (potential energy driven 
superconductivity).
The 2D pair-condensate can be characterized by a charge ordered state
with a "checkerboard" like pattern seen by scanning tunneling microscopy.
The drop of the $c$-axis dielectric screening can be the primary source of the condensation energy 
at optimal doping.
We find that Coulomb energy gain occurs along the $c$-axis, which is proportional to the measured condensation energy ($U_0$) and to $T_c$:
$E_c^{3D} \approx 2 (\xi_{ab}/a_0+1)^2 U_0 \approx k_B T_c$ and is due to inter-layer charge complementarity (charge asymmetry of the boson condensate)
where $\xi_{ab}$ is the coherence length of the condensate and $a_0 \approx 3.9 \AA$ is the in-plane lattice constant.
The static $c$-axis dielectric constant $\epsilon_c$ is calculated for various cuprates
and compared with the available experimental data. 

\end{abstract}
}


 Despite the enormous amount of effort spent in the last decades in resolving the fundamental physics behind
high temperature superconductivity (HTSC) the mechanism is still puzzling \cite{Hirsch}.
 It is more or less generally accepted now that the conventional electron-phonon pairing mechanism 
cannot explain cuprate superconductivity, because as high a transition temperature as
$164 K$ (the record $T_c$ up to now \cite{Xiong}) cannot be explained by the energy scale of lattice vibrations without leading to lattice instability \cite{Hirsch}.
It is, however, already well established that much of the physics related to HTSC
is in 2D nature,
one of the basic questions to be answered still in the future is whether HTSC is
a strictly 2D phenomenon or should be described by a 3D theory including all the 2D
features of superconductivity.

 The systematic dependence of the transition temperature $T_c$ on the $c$-axis
structure and, in particular, on the number of $CuO_2$ planes in multilayer blocks are 
strongly in favour of the 3D character of HTSC.
Variation of the thickness of ultrathin artificial HTSC compounds leads to the significant change
of $T_c$ supporting the importance of IL coupling in cuprates \cite{Li}.
 It has also been shown recently that the $c$-axis transport of the hole-content to the doping site in the charge reservoir in the normal state (NS)
of cuprates may be important in uderstanding HTSC 
\cite{Sule}.
Despite of the significant amount of published works which support the importance of the 3D character
of HTSC 2D theories rule the literature in the last years \cite{Hirsch}.
This is partly due to the failure of the inter-layer (IL) tunneling model of Anderson \cite{PWA,Basov} which was the
prominant 3D theory of HTSC.
Another reason could be that no sraigthforward dependence of $T_c$ can be seen on the IL distance
and on other 3D parameters of cuprates. Therefore the 3D character of HTSC is more complicated
and needs more theoretical and experimental work in the future.

  In this paper we propose a simple phenomenological model for explaining the 3D character of HTSC
in cuprates supported by calculations.
We would like to study the magnitude of direct Coulomb interaction
between charge ordered square superlattice layers as a possible source of pairing interaction.
Our intention is to understand HTSC within the context of an IL Coulomb-mediated mechanism.
 The IL charging energy we wish to calculate depends on the IL spacing ($d$), the IL dielectric
constant $\epsilon_c$, the hole content $p$ and the size of the superlattice.
A large body of experimental data are collected which support our model.
A consistent picture is emerged on the basis of the
careful analysis of this data set. 
Finally we calculate the static $c$-axis dielectric constant $\epsilon_c$ (that is normal to the $ab$-plane) for various cuprates
which are compared with the available experimental observations.

 {\em We propose to examine the following charge ordered superlattice model of pair condensation:}
 A pair of charge carriers ($2e$) can in principle be distributed over $2/0.16=12.5$ $CuO_2$ unit cells in a square
lattice layer if the $2e$ pair is composed of the optimal hole content $p_0 \approx 0.16/CuO_2$ at optimal doping \cite{Presland}.
However, allowing the phase separation of hole-anti-hole pairs, every second unit cell is occupied by
$-0.16e$ (anti-hole), and the rest is empty (holes, $+0.16e$), therefore we have $25$ unit cells for a condensed pair of charge 
carriers
 (FIG 1., that is the unit lattice of the pair condensate in the $5 \times 5$ supercell model).  
Therefore, $2 p_0 =-0.32e$ hole charge condenses to every second hole forming anti-hole sites.
The $25$ lattice sites provide 
the hole content of $25 \times 0.16e=4e$ and therefore $2e$ excess charge in the anti-hole sites.
The other $2e$ charge is used for neutralizing the holes at the anti-hole sites.
In other words a Cooper wave-function can be distributed on a $5 \times 5$ superlattice
with a node of $-0.16e$ in every anti-hole site.
The remarkable feature is that the size of the $5 \times 5$ condensate (four lattice spacings, $4 a_0 \approx 15.5 \AA$) is comparable with the measured small coherence
length $\xi_{ab}$ of single-layer cuprates ($\xi_{ab} \sim 10 \AA$ to $20 \AA$) \cite{Tinkham}.
$\xi_{ab}$ can directly be related to the characteristic size of the wave-pocket of the local Cooper pair (coherence area) \cite{Tinkham}.
The {\em charge ordered superlattice bilayer} (COSB) model can in principle be applied not only for $p_0 \approx 0.16e$ but also for the entire doping regime.

  
  IL Coulomb energy gain occurs only in that case when 
holes in one of the layers are in proximity with anti-holes in the other layer
(FIG 1, IL {\em electrostatic complementarity}, bilayer $5 \times 5$ ($4a_0 \times 4a_0$) model).
An important feature is then that
the boson condensate can be described by an IL {\em charge asymmetry}.
The IL coupling of the boson-boson pairs in the bilayer $5 \times 5$ model naturally suggests the
effective mass of charge carriers $m^{*} \approx 4 m_e$, as it was found by measurements \cite{krusin}.
The $5 \times 5$ model can be generalized to represent a $N \times N$ coherence
area where $N$ is the real space periodicity of the superlattice.
In the $N \times N$ superlattice model the hole and anti-hole partial charges at each $CuO_2$ lattice
sites are $q=\pm 4e/N^2$ \cite{Sule:condmat}. Important to note that the charge sume rule
holds for the characteristic bilayer with a coherence area $\sum_i^{N^2} q_i^{ahole}=4e$
where $q_i^{ahole}$ is the partial anti-hole charge at the $i$th anti-hole lattice site.
In other words a pair of a boson condensate can be accompanied within a COSB depicted in FIG 1.

 Important consequence of the model outlined in this paper is that the condensation
energy ($U_0$) of the SC state can be calculated.
We will show here that
within our model
the primary source of $U_0$ is the net energy gain in IL Coulomb energy occurs due to  
the asymmetrical distribution of the condensed hole-charge
in the adjacent layers (see FIG 1) below $T_c$.

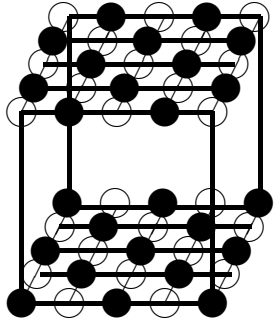
\begin{figure}

\setlength{\unitlength}{0.05in}
\begin{picture}(30,30)(-22,4)
\linethickness{0.55mm}
  \put(0,0){\line(1,2){5.0}}
  \put(20,0){\line(1,2){5.0}}
  \put(0,20){\line(1,2){5.0}}
  \put(20,20){\line(1,2){5.0}}
  \put(5,0){\line(1,2){5.0}}
  \put(10,0){\line(1,2){5.0}}
  \put(15,0){\line(1,2){5.0}}
  \put(5,20){\line(1,2){5.0}}
  \put(10,20){\line(1,2){5.0}}
  \put(15,20){\line(1,2){5.0}}
  \put(20,0){\line(0,1){20.0}}
  \put(0,0){\line(0,1){20.0}}
  \put(0,20){\line(1,0){20.0}}
  \put(0,0){\line(1,0){20.0}}
  \put(5,30){\line(1,0){20.0}}
  \put(5,10){\line(0,1){20.0}}
  \put(25,10){\line(0,1){20.0}}
  \put(5,10){\line(1,0){20.0}}
  \put(2,3){\line(1,0){20.0}}
  \put(2.5,5.5){\line(1,0){20.0}}
  \put(4,8.){\line(1,0){20.0}}
  \put(1.3,22.5){\line(1,0){20.0}}
  \put(2.3,25.0){\line(1,0){20.0}}
  \put(3.5,27.5){\line(1,0){20.0}}
  \put(0,0){\circle*{3}}
  \put(5,0){\circle{3}}
  \put(10,0){\circle*{3}}
  \put(15,0){\circle{3}}
  \put(20,0){\circle*{3}}
  \put(0,20){\circle{3}}
  \put(5,20){\circle*{3}}
  \put(10,20){\circle{3}}
  \put(15,20){\circle*{3}}
  \put(20,20){\circle{3}}
  \put(1.6,3){\circle{3}}
  \put(6.3,3){\circle*{3}}
  \put(11.5,3){\circle{3}}
  \put(16.5,3){\circle*{3}}
  \put(21.3,3){\circle{3}}
  \put(2.5,5.5){\circle*{3}}
  \put(7.5,5.5){\circle{3}}
  \put(12.5,5.5){\circle*{3}}
  \put(17.5,5.5){\circle{3}}
  \put(22.5,5.5){\circle*{3}}
  \put(4,8){\circle{3}}
  \put(8.6,8){\circle*{3}}
  \put(13.8,8){\circle{3}}
  \put(18.8,8){\circle*{3}}
  \put(23.8,8){\circle{3}}
  \put(4.8,10.5){\circle*{3}}
  \put(9.8,10.5){\circle{3}}
  \put(14.8,10.5){\circle*{3}}
  \put(19.8,10.5){\circle{3}}
  \put(24.8,10.5){\circle*{3}}
  \put(1.3,22.5){\circle*{3}}
  \put(6.3,22.5){\circle{3}}
  \put(10.8,22.5){\circle*{3}}
  \put(16.4,22.5){\circle{3}}
  \put(21.4,22.5){\circle*{3}}
  \put(2.3,25.0){\circle{3}}
  \put(7.3,25.0){\circle*{3}}
  \put(12.3,25.0){\circle{3}}
  \put(17.3,25.0){\circle*{3}}
  \put(22.2,25){\circle{3}}
  \put(3.3,27.5){\circle*{3}}
  \put(8.5,27.5){\circle{3}}
  \put(13.2,27.5){\circle*{3}}
  \put(18.3,27.5){\circle{3}}
  \put(23,27.5){\circle*{3}}
  \put(4.4,30.0){\circle{3}}
  \put(9.4,30.0){\circle*{3}}
  \put(14.4,30.0){\circle{3}}
  \put(19.4,30.0){\circle*{3}}
  \put(24.4,30.0){\circle{3}}
\end{picture}

\vspace{1cm}
\caption{\small The charge ordered state of the hole-anti-hole condensate
in the $4a_0 \times 4a_0$ charge ordered bilayer superlattice model. Note the charge asymmetry between the adjacent layers. The bilayer can accomodate a pair of boson condensate ($4e$).
The inter-layer charge complementarity of the charge ordered state is crucial for getting inter-layer Coulomb
energy gain.
}

\end{figure}

Per definition $U_0$ is the free energy difference of the normal state
and the superconducting state,
\be
U_0 \approx \Delta E^{ab} + \Delta E_c^{IL}.
\ee
Tha $ab$-plane contribution to the SC condensation energy $\Delta E^{ab}$ is expected to be
negligible at optimal doping.
That is because our Coulomb energy calculations indicate that the $ab$-plane Coulomb energy in the NS and in the SC state are nearly equal
and therefore the $ab$ Coulomb energy contribution to $U_0$  $\Delta E_c^{ab} \approx 0$ 
and also the kinetic energy of the pair-condensate $E_{kin}^{ab}$ is much smaller
by several order of magnitude then IL coupling.
The effective Coulomb interaction between spinless point-charges
may be approximated by the expression $V_{eff}({\bf {r}})=e^2/(4 \pi \epsilon_0 \epsilon_c {
\bf {r}})$, where $\epsilon_c$ takes into account phenomenologically the dielectric screening
effect of the IL dielectric medium and confined hole charge \cite{Sule:condmat}.
For instance we get for the prototypical $5 \times 5$ lattice $\sim -1.2 \times 10^{-20}$ J $ab$-Coulomb
energy both for the NS and for the SC COS using a simple point charge model for calculating
the Coulomb energy. The details of this calculation can can found in ref. \cite{Sule:condmat}.
The calculated $ab$-kinetic energy of the $5 \times 5$ codensate is $\sim 1.1 \times 10^{-29}$ J 
if we use the very simple formula ("electron in a box")
\be
T_{kin}^{ab} \approx \frac{\hbar^2}{2m^*} \frac{2e}{\xi_{ab}^2 \xi_c},
\ee
where $\xi_c$ is the $c$-axis coherence length.
The IL Coulomb interaction (attraction) $E_c^{IL} \approx -1.3 \times 10^{-21}$ J for the $5 \times 5$ bilayer. 
It must be emphasized, however, that away from optimal doping
the Coulomb energy of the condensate might affect the magnitude of the condensation energy.
The abrupt jump of the measured condensation energy $U_0$ seen in the slightly overdoped regime
\cite{TallonLoram} can be attributed to the increased $\Delta E^{ab}$ contribution to $U_0$. 

  In this article we restrict our study, however, to optimal doping and the study of the doping dependence of
$U_0$ is planned in the near future.
The main contribution to $U_0$ is then the IL energy gain $\Delta E_c^{IL}$ 
in the SC state can be given as follows,
\be
  U_0 \approx \vert E_c^{IL,NS}-E_c^{IL,SC} \vert.
\label{gain}
\ee
where $E_c^{IL,NS}$ and $E_c^{IL,SC}$ are the IL Coulomb energy in the NS and in SC state.
We can further simplify Eq.~(\ref{gain}) if
the NS contribution to Eq.~(\ref{gain}) is
$E_c^{IL,NS} \approx 0$, which holds if IL coupling is screened effectively in the NS (large density of the hole
content in the IL space, large $\epsilon_c(NS)$).
Again, if we assume $\epsilon_c \approx 33$, we get $E_c^{IL,NS} \approx +2 \times 10^{-18}$ J ($12.3$ eV)
for a charge ordered $5 \times 5$ lattice
which would be an extraordinarily large value for the IL Coulomb repulsion.
Assuming, however, $\epsilon_c \approx 10000$, we get the more realistic Coulomb repulsion of $E_c^{IL,NS} \approx +10^{-22}$ J.
Indeed there are measurements for cuprates which indicate large $\epsilon_c$ in the NS
in the range of $10^3$ to $10^5$ \cite{Cao}. 
Extraordinarily large dielectric constant has also been found recently
in perovskite materials \cite{Homes}.
Anyhow the "relaxation" of the huge IL Coulomb repulsion in hole-doped cuprates in the NS can not easily be
understood without the consideration of a large $\epsilon_c$.
Nevertheless experimental measurements and theoretical speculations suggest that
$\epsilon_c$ in cuprates and in other materials with perovskite structure is strongly temperature and
doping dependent \cite{Kitano}. 
The sharp decrease of $\epsilon_c$ below $T_c$ should also be explained within our bilayer
model by the pair-condensation of the hole-content to the sheets.
This is what can be seen in the $c$-axis optical spectra of cuprates \cite{Basov,Molegraaf}.
The lack of the $c$-axis plasma edge in the NS is an obvious experimental evidence of the
strong temperature dependence of $\epsilon_c$.
  Then we have
\be
2 (n+1) N^2 U_0 \approx E_c^{IL,SC},
\label{gain_sc}
\ee
where $E_c^{IL,SC}$ is the Coulomb energy gain in the SC state.
$U_0$ is the experimental condensation energy given per unit cell. 
Eq.~(\ref{gain_sc}) is generalized for multilayer cuprates introducing
$n$.
For single layer cuprates $n=0$, for bilayers $n=1$, etc.

\begin{table}
\caption[]
{The calculated coherence length of the pair condensate
given in $a_0$ using the experimental condensation energies
of various cuprates and Eq.~(\ref{N}) at optimal doping.
}
{\scriptsize
\begin{tabular}{cccccc}
 & $T_c$ (K) & $k_B T_c$ (meV) & $U_0$ ($\mu eV/u.c.$)  & $\xi_{ab}^{calc} (a_0)$ & $\xi_{ab}^{exp} (a_0)$ \\ 
\hline
  Bi2201 & 20 & 1.6 & $10^a$  &  $\sim 8$ &       \\
 LSCO   & 39  & 2.5 & $21^b$ &  $\sim 7$ &  $5-8^c$  
 \\
 Tl2201 & 85 & 7 & $100 \pm 20^d$ & $\sim 5$ & \\
 Hg1201 & 95 & 7.8 & $80-107^e$ &  $\sim 5$ & $5^f$ \\
  YBCO  & 92 & 7.5 & $110^g$ &  $\sim 3$ & $ 3-4^h $ \\
  Bi2212 & 89 & 7.3 & $95^g$ &  $\sim 3-4$ & $4-6^i$\\
\end{tabular}}
{\scriptsize
$a_0 \approx 3.88 \AA$,
$U_0$ is the measured condensation energy of various cuprates
in $\mu$ eV per unit cell at optimal doping.
$^a$ from \cite{MarelPC},
$^b$ $U_0 \approx 2$ J/mol from \cite{Loram,Momono},
$^c$ from \cite{Tinkham},
$^d$ \cite{Tsvetkov},
$^e$ $U_0 \approx 12-16$ mJ/g from \cite{Billon,Kirtley} and $\xi_{ab}$ from \cite{Thompson},
$^d$ $U_0 \approx 11$ J/mol from \cite{TallonLoram},
$^e$ $U_0 \approx 10$ J/mol from \cite{TallonLoram},
$^f$ from \cite{Thompson},
$^g$ from \cite{TallonLoram},
$^i$ from recent measurements of Wang {\em et al.}, $\xi_{ab} \approx 23 \AA (\sim 5-6 a_0)$ \cite{Wang_sci},from STM images of ref. \cite{Hoffman} $\xi_{ab}\approx 4a_0$,
$^h$ from \cite{Tinkham},
$\xi_{ab}^{calc}$ is calculated according to Eq.~(\ref{N}) and is also given in Table ~\ref{tab1} and $\xi_{ab}^{exp}$ is
the measured in-plane coherence length given in $a_0 \approx 3.9 \AA$.
The notations are as follows for the compounds:
Bi2201 is $Bi_2Sr_2CuO_{6+\delta}$,
LSCO ($La_{1.85}Sr_{0.15}CuO_4$),  
Tl2201 ($Tl_2Ba_2CuO_6$), 
Hg1201 ($HgBa_2CuO_{4+\delta}$),  
 YBCO ($YBa_2CuO_7$) and
 Bi2212 is $Bi_2Sr_2CaCu_2O_{8+\delta}$.
}
\label{tab1}
\end{table}


Inter-layer Coulomb coupling can be a natural source of the condensation
energy and of HTSC.
Checkerboard-like charge pattern (COS) seen experimentally \cite{Hoffman,Lake} directly leads to IL energy gain and
to potential energy driven condensation energy if the hole-anti hole
charge pattern is asymmetrycally condensed to the adjacent layers (FIG 1).
Assuming that thermical equilibrium occurs at $T_c$ for the competing
charge ordered phases of the NS and the SC state the following equation
for the condensation energy can be formulated using Eq.~(\ref{gain_sc}),


\begin{figure}[hbtp]
\begin{center}
\includegraphics*[height=4.5cm,width=6.5cm]{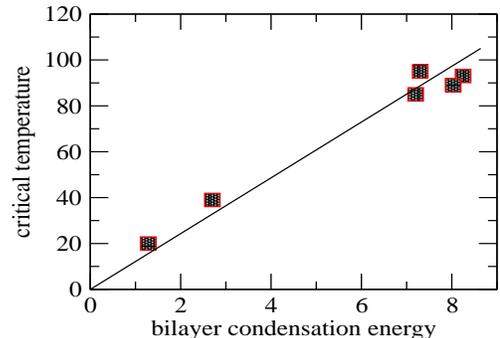}
\caption[]{
The critical temperature (K) at optimal doping as a function of the bilayer condensation energy
($2 (n+1) N^2 U_0$, meV) using Eq.~(\ref{kbtc}).
The straight line is a linear fit to the data. The slope of the linear fit
is $k_B$ which is a strong evidence of Eq.~(\ref{kbtc}).
}
\label{cond_tc}
\end{center}
\end{figure}


\be
2 (n+1) N^2  U_0 \approx 2 (n+1) \biggm[ \frac{\xi_{ab}}{a_0}+1 \biggm]^2 U_0 \approx k_B T_c,
\label{kbtc}
\ee
where $k_B$ is the Boltzmann constant.

At a first look this formula seems to be unusual because of the dependence
of the condensation energy on $T_c$. The available measurements of the
condensation energy on various cuprates show no correlation of $U_0$
with the critical temperature.
  One of the important goals of this paper, however, to show that correlation can indeed
be found with $T_c$ if $T_c$ is plotted against $2 (n+1) N^2 U_0$ (the bilayer condensation energy, FIG 2).
 In other words {\em the condensation energy of the COSB 
system shows correlation with $T_c$}.
 In order to test the validity of Eq.~(\ref{kbtc}) we estimate the 
coherence length of the pair condensate using Eqs.~(\ref{kbtc})
and using only experimental data,
\be
\xi_{ab} \approx a_0 \biggm[ \sqrt{\frac{k_B T_c}{2 (n+1) U_0}}-1 \biggm]
\label{N}
\ee
The results are given in Table~\ref{tab1} as $\xi_{ab}^{calc}$ and compared with the available
measured $\xi_{ab}^{exp}$. The agreement is excellent which strongly suggests that Eq.~(\ref{N})
should also work for other cuprates.
 The validity of Eq.~(\ref{kbtc}) is clearly clarified in FIG~\ref{cond_tc} using mostly experimental
information for $\xi_{ab}$, $T_c$ and $U_0$ (values are given in Table I.).
For Tl2201 no measured $\xi_{ab}$ is found in the literature, therefore the estimated 
$\xi_{ab}$ is used (given in Table I.). We will show that 
using this estimated value of $N=6 (\xi_{ab} \approx 5 a_0)$ we predict $\epsilon_c$ in nice agreement with
optical measurements for Tl2201.
{\em Remarkable feature of FIG~\ref{cond_tc} is that the slope of the linear fit to the
measured data points
is precisely $k_B$ which supports Eq.~(\ref{kbtc}).
}
The underlying physics of HTSC seems to be reflected by Eq.~(\ref{kbtc}):
the equation couples the basic observable quantities $T_c, U_0$ and $\xi_{ab}$.
We predict for the multilayer Hg-cuprates, Hg1212 ($T_c=126$ K) and Hg1223 ($T_c=135$ K) the condensation energies
$146$ and $117$ $\mu$eV/u.c. ($\xi_{ab} \approx 4 a_0$ \cite{Thompson}), respectively
using Eq.~(\ref{kbtc}).

 The expression Eq.~(\ref{kbtc}) leads to the very simple formula for the critical temperature
using Eq.~(\ref{gain_sc}) and a simple Coulomb expression for the IL coupling energy $E_c^{IL,SC}$ \cite{Sule:condmat}
($T_c \approx k_B^{-1} E_c^{IL,SC}$)
\be
T_c (N,d,\epsilon_c) \approx \frac{e^2}{4 \pi \epsilon_0 \epsilon_c k_B} \sum_{n=1}^{2}\sum_{m=2}^{N_l} \sum_{ij}^{N^2} \frac{q_i^{(n)} q_j^{(m)}}{r_{ij}^{(n,m)}}
\label{kbtc2}
\ee
where $r_{ij}^{(n,m)}$ is the inter-point charge distance and $r_{ij}^{(n,m)} \ge d_{IL}$, where $d_{IL}$ is the IL distance ($CuO_2$ plane to 
plane, $i \ne j$). $n,m$ represent
layer indexes ($n \ne m$). 
First the summation goes within the bilayer up to $N^2$ then the IL Coulomb interaction of the
basal bilayer are calculated with other layers
along the $c$-axis in both direction ($n=1,2$) \cite{Sule:condmat}.
$N_l$ is the number of layers along the $c$-axis. When $N_l \rightarrow \infty$, bulk $T_c$ is calculated.
$T_c$ can also be calculated for thin films when $N_l$ is finite.
$\epsilon_c$ can also be derived
\be
\epsilon_c \approx \frac{e^2}{4 \pi \epsilon_0 k_B T_c} \sum_{n=1}^{2}\sum_{m=2}^{N_l} \sum_{ij}^{N^2} \frac{q_i^{(n)} q_j^{(m)}}{r_{ij}^{(n,m)}}
\label{epsc4}
\ee
where a $c$-axis average of $\epsilon_c$ is computed when $N_l \rightarrow \infty$.

\begin{table}
\caption[]
{The calculated dielectric constant $\epsilon_c$ using Eq.~(\ref{epsc4})
in various cuprates at the calculated coherence length $\xi_{ab}$ of the charge ordered state
given in Table I.
}
{\scriptsize
\begin{tabular}{lccccc}
 & $d (\AA)$ & $T_c (K)$ & $\xi_{ab}^{calc} (a_0)$ & $\epsilon_c$ & $\epsilon_c^{exp}$  \\ \hline
 Bi2201 &    12.2 & 20   &   8  &  9.9 & 12$^a$ \\
 LSCO &      6.65 &  39 & 7 &  11.3 & $23 \pm 3, 13.5^b$  \\
 Hg1201    &     9.5   & 95  & 5 & 6.5 &    \\
 Tl2201 &    11.6 &  85 &  5 & 13.0 & 11.3$^c$ \\
  YBCO  &   8.5  & 93  &  3 & 19.4 & 23.6$^d$ \\ 
\end{tabular}}
{\scriptsize
where $\xi_{ab}^{calc}$ is the estimated in-plane coherence length given in $a_0 \approx 3.9 \AA$ unit.
$d$ is the $CuO_2$ plane to plane inter-layer distance in $\AA$, $T_c$ is the experime
ntal critical temperature.
$\epsilon_c$ is from Eq.~(\ref{epsc4}).
$\epsilon_c^{exp}$ are the measured values obtained from the following references: 
$^a$ \cite{Boris}, 
$^b$ \cite{epsilonc}, or from reflectivity measurements \cite{Tsvetkov},
$\omega_p \approx 55 cm^{-1}$ \cite{Sarma}, $\lambda_c \approx 3 \mu$m \cite{Kirtley},
$^c$ from \cite{Tsvetkov}, 
$^d$ from reflectivity measurements: $\omega_p \approx 60 cm^{-1}$ \cite{Sarma}, $\lambda_c \approx 0.9 \mu$m \cite{Kitano2}.
}
\label{tab2}
\end{table}

The calculation of the $c$-axis dielectric constants $\epsilon_c$ might provide
further evidences for Eq.~(\ref{kbtc}) when compared with the measured values \cite{Kitano,epsilonc}.
In Table ~\ref{tab2} we have calculated the static dielectric function $\epsilon_c$ using
Eq.~(\ref{epsc4}) and compared with the experimental impedance measurements \cite{Cao,epsilonc}.
$\epsilon_c$ can also be extracted from the $c$-axis optical measurements using the
relation \cite{Tsvetkov}
$\epsilon_c(\omega) =\epsilon_c(\infty)-c^2/(\omega_p^2 \lambda_c^2)$, 
where $\epsilon_{\infty}$ and $\omega_p$ are the high-frequency dielectric constant and the plasma frequency, respectively \cite{Tamasaku}. $c$ and $\lambda_c$ are the speed of light
and the $c$-axis penetration depth. 
At zero crossing $\epsilon_c(\omega)=0$ and $\omega_p=c/(\lambda_c \epsilon_c^{1/2}(\infty))$.
Using this relation we predict for the single layer Hg1201 the low plasma frequency of
$\omega_p \approx 8$ cm$^{-1}$ using $\epsilon_c=\epsilon_c(\infty) \approx 6.5$ (Table II) and $\lambda_c \approx
8$ $\mu$m \cite{Kirtley}. 

  If the physical picture derived from our model is correct, it should be a guide for 
further experimental studies aiming to improve SC in cuprates or in other materials.
This can be done by tuning the IL distance and $\epsilon_c$ in these materials.
The application of this model to other class of HTSC materials, such as fullerides or
$MgB_2$ is expected to be also effective.


{\small
It is a privilige to thank M. Menyh\'ard for the continous support.
I greatly indebted to E. Sherman for reading the manuscript carefully
and for the helpful informations.
This work is supported by the OTKA grant F037710
from the Hungarian Academy of Sciences}
\\


\end{document}